\documentclass[conference]{IEEEtran}
\IEEEoverridecommandlockouts

\usepackage{cite}
\usepackage{amsmath,amssymb,amsfonts}
\usepackage{algorithmic}
\usepackage{graphicx}
\usepackage{textcomp}
\usepackage{xcolor}
\usepackage{verbatim}
\usepackage{array}
\usepackage{url}
\usepackage{subfigure}
\usepackage{comment}
\usepackage{esdiff}
\usepackage{multirow}
\usepackage{multicol}
\usepackage{tablefootnote}
\usepackage[flushleft]{threeparttable}
\usepackage[symbol]{footmisc}
\def\BibTeX{{\rm B\kern-.05em{\sc i\kern-.025em b}\kern-.08em
    T\kern-.1667em\lower.7ex\hbox{E}\kern-.125emX}}
    
\renewcommand{\baselinestretch}{0.96}

\begin{document}

\title{A High Performance and Robust FIFO Synchronizer-Interface for Crossing Clock Domains in SFQ Logic}

\author{\IEEEauthorblockN{Anonymous Authors}
\IEEEauthorblockA{Paper under Blind Review}
}


\author{\IEEEauthorblockN{
Gourav Datta$^{\ddagger}$, 
Shidie Lin$^{\ddagger}$, Peter A.~Beerel}
\thanks{$^{\ddagger}$Authors have equal contribution.}
\IEEEauthorblockA{\textit{Ming Hsieh Department of Electrical and Computer Engineering} \\
\textit{University of Southern California}\\
Los Angeles, California 90089, USA \\
\{gdatta, shidieli, pabeerel\}@usc.edu}
}

\maketitle

\begin{abstract}

Digital single-flux quantum (SFQ) technology promises to meet the demands of ultra low power and high speed computing needed for future exascale supercomputing platforms. However, clocking SFQ logic circuits remains a challenge due to the presence of a large number of on-chip clock sinks, and hence, decomposing large designs into multiple independent clock domains similar to CMOS, have been proposed. However, such clock domains demand efficient synchronizing First-in-first-out (FIFO) buffers and robust interfaces to safely transfer data from one clock domain to another. In this brief, we propose such a FIFO synchronizer and clock-domain crossing interface for both uni and bi-directional data transfer 
without any significant degradation of the clock frequency. Our proposal 
scales to complex gate-level pipelined SFQ logic cores while demonstrating extremely low Bit Error Rate (BER) and is unaffected by noise, given current SFQ lithography feature sizes.

\end{abstract}

\section{Introduction}\label{sec:intro}

As we face the fundamental limits of physical scaling dictated by Moores' law, single flux quantum (SFQ) \cite{likharev1991} has emerged as a promising beyond-CMOS logic technology, thanks to its switching energy per bit of ${\sim}10^{-19}$ J \cite{switch_energy1} and the potential to support clock frequencies up to $770$ GHz \cite{high_freq_sfq}. Motivated by its' promise, many researchers have successfully demonstrated cryogenic arithmetic logic units \cite{alu_isec} and low-complexity microprocessors \cite{core_1b,COREe4, FLUX2001}. Recently, several energy-efficient variants of SFQ technologies have also been explored  \cite{mukhanov2011, mukhanov2, low_power_sfq, low_power_sfq1}. 
However, the potential of three orders of magnitude lower power consumption (in the case of non-resistive bias networks \cite{mukhanov2011}) at an order of magnitude higher frequency \cite{likharev1991}, has still not been realized, primarily due to i) high process variations and non-idealities \cite{likharev1991, parameters1995}, and ii) the lack of a three-terminal controllable switch element.

In particular, the gate-level pipelining and ultra-high clock frequencies associated with SFQ logic makes low-skew clock distribution extremely challenging \cite{timingSpringer}. As a result, a $1$ THz device was forced to function at a disastrous $20$ GHz frequency \cite{bunyk50}. Previous work \cite{gdatta_tas} addressed this clocking challenge by decomposing the SFQ design into several independently clocked blocks, i.e., into multiple clock domains, similar to how heterogeneous CMOS designs are managed. 
Moreover, increasing integration densities in SFQ logic \cite{sfq_scalable} can drive SFQ circuit designers to implement increasing numbers of on-chip clock domains. 
However, in the traditional design approach, circuitry within each clock domain is designed using a dedicated clock. Since these clock domains have no phase relationship, static timing constraints cannot be created to guarantee safe data transfer. Hence, the setup time of flip-flops (FFs) at the boundary of these domains may be violated. They can thus exhibit metastability \cite{ginosar_sync} and high BER.

As timing constraints cannot be guaranteed between clock domains, safe/robust communication between them sometimes takes place either at a rate slower than the system clock (e.g. one transfer for every two cycles of the clock) or with some kind of mixed asynchronous design \cite{ginosar_sync,abdelhadi2019bundled}. While synchronizers for SFQ logic have been previously proposed  \cite{gdatta_tas}, this paper is the first to propose a robust, high performance synchronizer that has flexible and robust read/write 
interfaces that support high-throughput bi-directional communication between asynchronous clock domains. 

The remainder of the paper is organized as follows. Section \ref{sec:background} provides related background on SFQ, including descriptions of the SFQ crossing clock domains and First-in-first-out (FIFO) synchronizers, and identifies the key bottlenecks in previous synchronizer designs \cite{gdatta_tas}. Section \ref{sec:proposed_design} presents two improvements over \cite{gdatta_tas} and the associated interface for uni-directional communication. Section \ref{sec:proposed_interface} proposes a custom interface design to enable bi-directional communication between two clock domains. 
Section \ref{sec:simulation_setup} performs JSIM \cite{jsim} simulations to demonstrate the efficacy of the proposed synchronizer-interface. Finally some conclusions are given in Section \ref{sec:conc}. 

\begin{figure}[!t]
\centering {
\includegraphics[width=9cm]{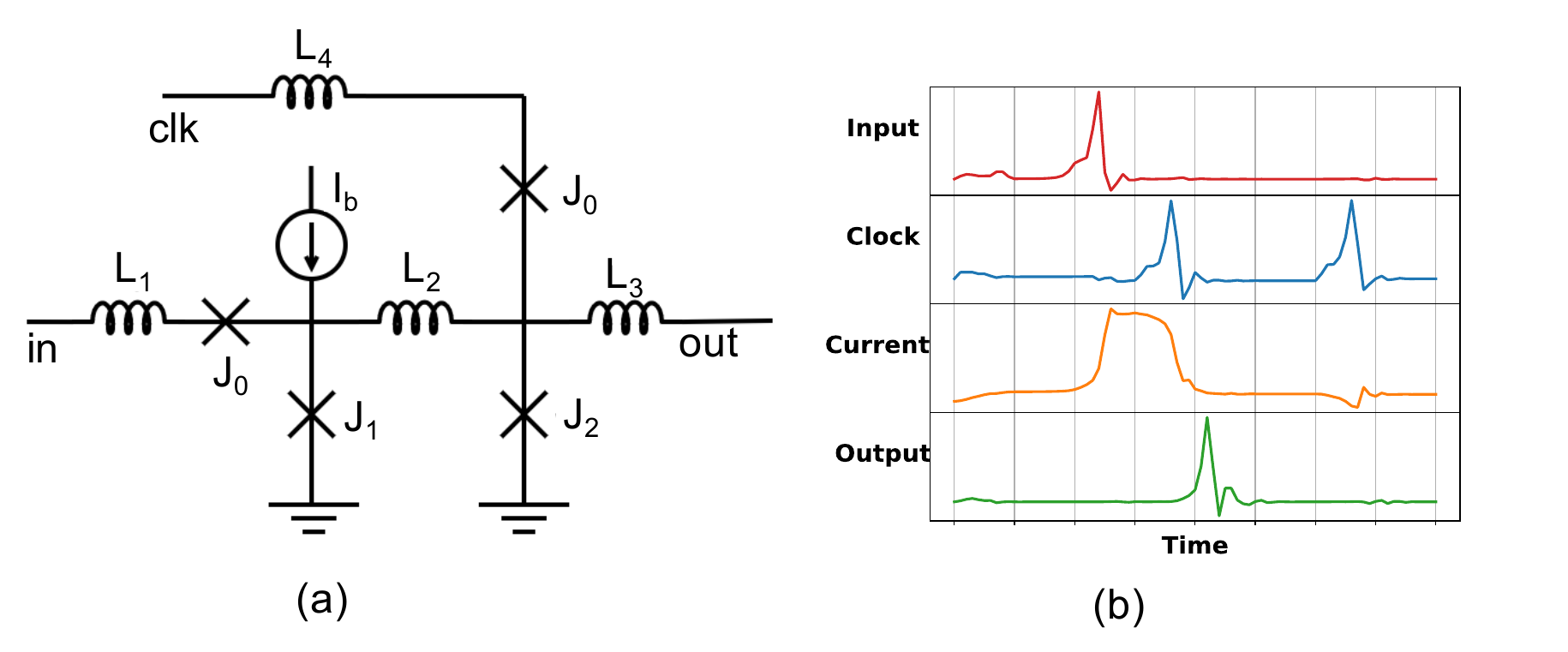}
\vspace*{-0.7cm}
\caption{(a) Schematic of a DFF (b) Simulation result of a DFF in SFQ 
}
\label{fig:dff}
}
\vspace*{-0.5cm}
\end{figure}

\section{Background \& Related Work}\label{sec:background}

\subsection{SFQ Logic}

\begin{figure}[!t]
\centering {
\includegraphics[width=8.8cm]{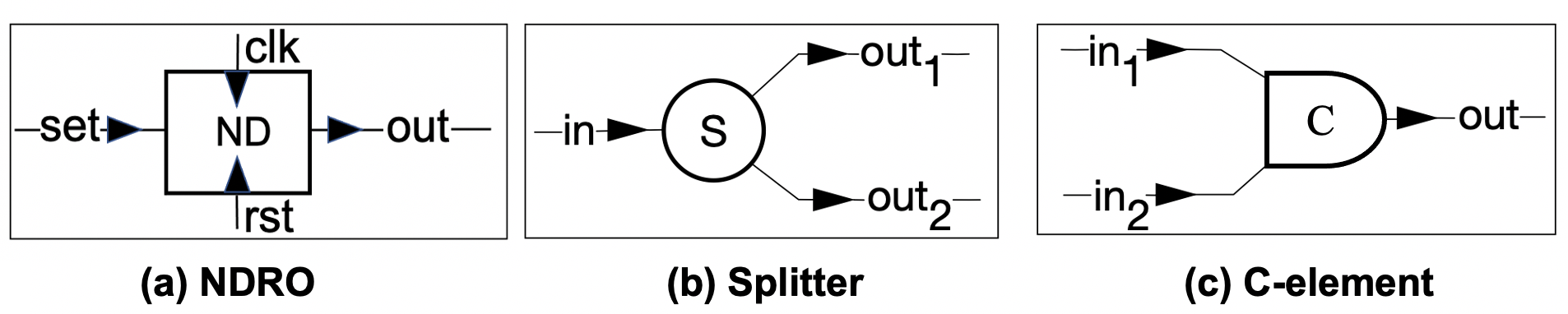}
\caption{Some basic SFQ cells' symbols 
}
\label{fig:sfq_cells}
}
\vspace{-.6cm}
\end{figure}

Unlike in CMOS, in SFQ technology, binary information is represented by very short (picosecond) voltage pulses $V(t)$ of quantized area, corresponding to transition of a single flux quantum, $\phi_{0}=\int V(t)dt=\frac{h}{2e}=2.03$ $mV.ps$.
These SFQ pulses can be quite naturally generated, reproduced, amplified, memorized, and processed by elementary cells comprising overdamped Josephson junctions (JJs) \cite{likharev1991}. 
In particular, the DC superconducting quantum interference device (SQUID) is the fundamental memory element that is used to store SFQ pulses \cite{bunyk50} and, to explain its use, we illustrate a SFQ D flip-flop (DFF) along with representative simulation waveforms in Figs. \ref{fig:dff}(a) and \ref{fig:dff}(b) respectively. 
A variant of the SFQ DFF which allows non-destructive readout (NDRO) of its' contents has also been demonstrated \cite{likharev1991} (see Fig. \ref{fig:sfq_cells}(a)). The SFQ DFF responds to its inputs by changing its internal loop current state, whereas the gate output will change only after the clock pulse arrives, which will reset the gate’s internal state and generate the correct gate output value. 

However, there are some SFQ cells which do not need a clock input. 
Unlike CMOS, in SFQ logic, a splitter, shown in Fig. \ref{fig:sfq_cells}(b), must be used in order to fan out a source signal to multiple destinations. Fig. \ref{fig:sfq_cells}(c) shows a 
C-element which produces an output SFQ pulse as soon as both its' inputs have been fed by such pulses, and follows a finite state machine (FSM) as described in \cite{likharev1991}. Its dotted counterpart uses a tweak in the bias current distribution circuit that results in the junction being initialized in a different state of the FSM. 
These asynchronous cells are the gate-level building blocks of our FIFO synchronizer and interface discussed in Section \ref{sec:proposed_design} and more details about SFQ logic can be found in \cite{likharev1991}.
 

\subsection{Crossing Clock Domains (CDC) in SFQ}



Clocking SFQ circuits is challenging because their deeply-pipelined nature makes them more sensitive to setup violations that can increase clock-to-Q delays than their CMOS counterparts. The high degrees of variability further exacerbates the ultra-high-speed low-skew clock distribution of large-scale SFQ circuits. In order to mitigate this problem, multiple independent SFQ clock domains can be used \cite{gdatta_tas}, which require robust design of reliable clock domain crossing (CDC) circuits. 
Although previous work \cite{gdatta_isec, gdatta_tas} presented a multi-flip-flop SFQ FIFO synchronizer, that simulations show delivers over a $1000$x reduction in BER at $30$ GHz, it did not propose an interface that enables bi-directional communication between clock domains. 
Moreover, \cite{gdatta_tas} also assumed the write side interface pushes data into the FIFO every clock cycle, which may not be true for practical applications. 

\section{Uni-directional Communication}\label{sec:proposed_design}

\begin{figure*}
\centering {
\includegraphics[width=0.8\textwidth]{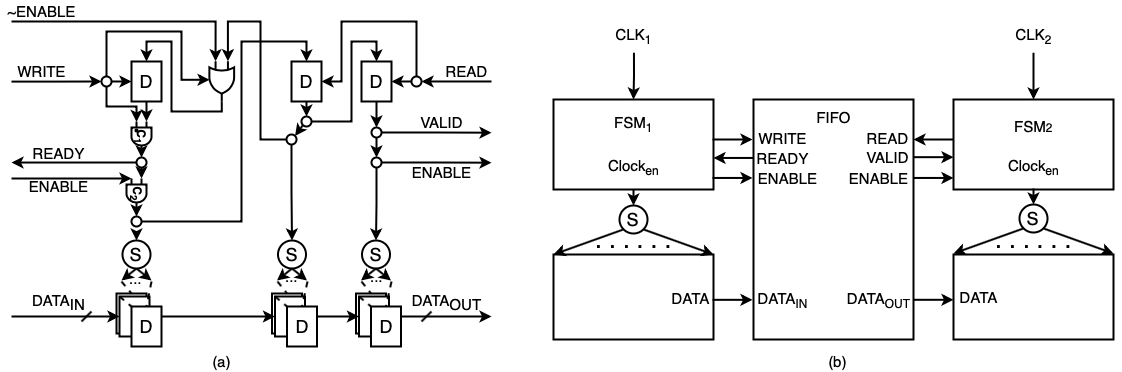}
\caption{(a) Proposed FIFO synchronizer and (b) associated Read and Write interfaces 
}
\vspace{-.1cm}
\label{fig:proposed_fifo}
}
\vspace{-.1cm}
\end{figure*}

In this section, we present our proposed synchronizer and interface, illustrating uni-directional communication between two unrelated clock domains. Our FIFO is shown in Fig. \ref{fig:proposed_fifo}(a).
\footnote{We show a 3-stage synchronizer, however, the optimal number of stages actually depends on throughput and burstiness constraints \cite{gdatta_isec}.} 

\subsection{Proposed Synchronizer}

The \textit{first improvement} in our proposed synchronizer over \cite{gdatta_tas,gdatta_isec} is that the newly proposed synchronizer accepts an additional enable signal from the write side notifying an intent to write data to the FIFO. Thus our proposed FIFO does not need to receive data every clock cycle, providing more flexibility to the designer. Moreover, this FIFO has a ready signal that is asserted when the FIFO has an empty pipeline stage in which it can accept new data. The cost of this flexibility, however, is that there is a two clock cycle delay between the enable and ready signal, and hence, we can write new data into the FIFO at most every other clock cycle.

In particular, after the ready and enable signals are asserted, the lower C-element $C_2$ produces a clock pulse that reads the new data into the FIFOs datapath and helps move this data forward through the FIFO with successive pulses on the read clock. 
The ready signal also sends out a pulse when the ${\sim}$enable is active and the FIFO can still accept new data. The combination of these two distinct situations is implemented using the OR gate in the FIFO control path.
Note that the enable and ${\sim}$enable signals are mutually exclusive, and should arrive in the same clock cycle as new data is available.


Since all SFQ logic gates are clocked, the OR gate, followed by the DFF, function similar to a two-flop synchronizer. Should the output of the synchronization OR gate become metastable, it still needs to propagate through the DFF next to it, before its value is used by the write side interface. The extra time provided by the additional synchronization DFF increases the probability that the metastable value will resolve, and is the \textit{second improvement} in our design. The result is that the FIFO has low BER in the write side, as detailed in Section \ref{sec:simulation_setup}. On the read side, as proposed in \cite{gdatta_tas}, the data and valid signal might come out in different clock cycles if the valid signal arrives late, thereby corrupting the data transfer. Therefore, inspired by \cite{gdatta_tas, dattaiscas2021}, we employ two back-to-back DFFs to drive the read interface and make the probability of a late valid signal negligible. 

Although the above description assumes noiseless synchronizers, a real system has noise and the resolution to $0$ or $1$ of a DFF synchronizer near its metastable point is a stochastic process. 
Fortunately, \cite{gdatta_tcas} proves that the presence of noise has almost no effect on the BER of a two-flop SFQ synchronizer due to metastability.
Since our design involves two sequential clocked logic gates (OR and DFF), the same conclusion holds true in this design.

\subsection{Interface}

Upon reset, the write side of the FIFO interface assumes the FIFO is empty and that it can start operating by sending in data to the FIFO by asserting the enable signal. The output of the synchronizing DFF in the control path, along with the split write clock pulse, trigger the FIFO C-element $C_1$ and its split output produces the ready pulse, informing the write interface that it can accept new (enabled) data.
Once consumed data propagates to the read side of the FIFO, a read clock pulse will produce a valid pulse, and the associated data can be read out. 
Unlike the write side, we can read out one data token every clock cycle, provided the read clock frequency abides by the timing constraints detailed in Sec \ref{sec:proposed_interface}-B.
In principle, generic finite state machines (FSMs), as shown in Fig. \ref{fig:proposed_fifo}(b), can manage the data transfer to/from our proposed FIFO protocol.

\begin{figure}[!t]
\centering {
\includegraphics[width=9cm]{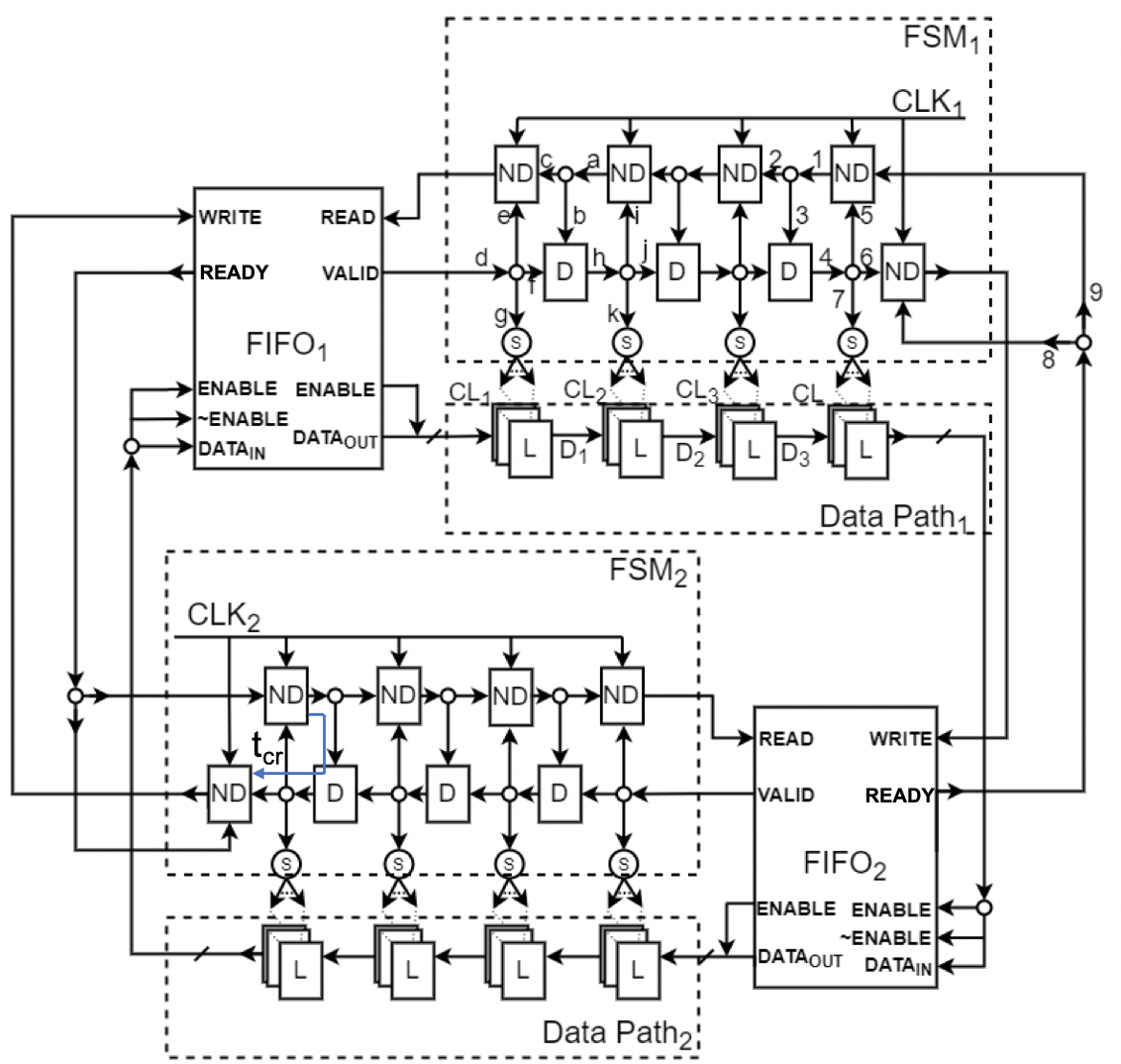}
\vspace{-0.6cm}
\caption{Proposed interface for an 8-stage circular shift register that spans two asynchronous clock domains with bi-directional data transfer. 
}
\label{fig:loop_circuit}
}
\vspace{-.4cm}
\end{figure}

\section{Bi-directional Communication}\label{sec:proposed_interface}


\begin{figure*}
\centering {
\includegraphics[width=0.95\textwidth]{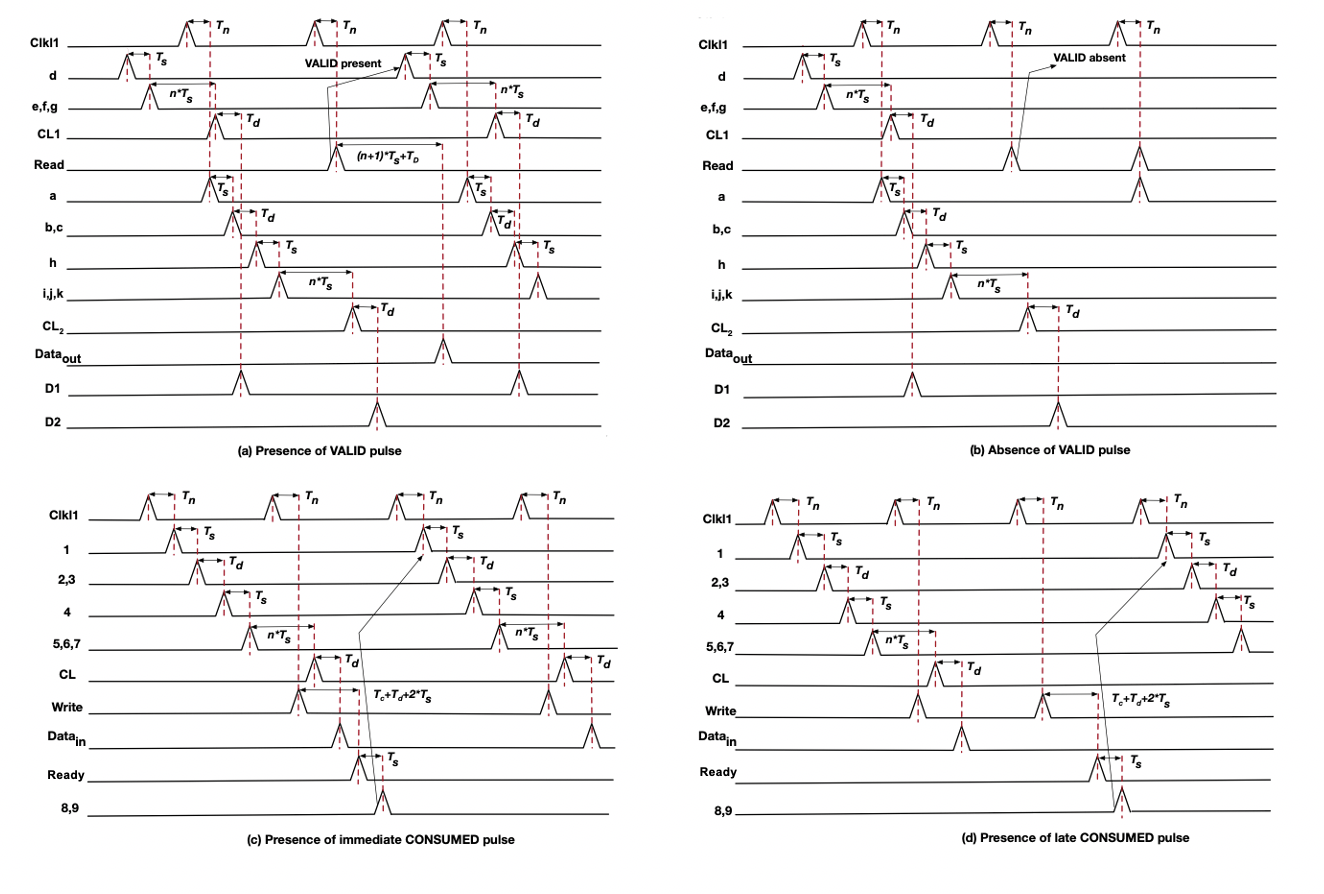}
\vspace{-0.5cm}
\caption{Timing diagram of the dataflow for 
the proposed circuit shown in Fig. \ref{fig:loop_circuit}.
The first two cases illustrate the read side of the FIFO$_1$ with and without the presence of a valid pulse and second two cases illustrate the corresponding behavior on write side of FIFO$_2$. $T_n$, $T_s$, $T_d$, $T_c$, and $n$ represent the NDRO clock-to-Q delay, splitter delay, DFF clock-to-Q delay, C-element delay, and the bit-width of the data respectively. 
}
\label{fig:timing_diagram}
}
\end{figure*}

Whenever incoming or outgoing FIFO data is blocked, the interface logic must stall to avoid losing new data or overwriting old data, respectively. Stalling is typically implemented by gating the associated interface clock. However, the gate-level clocking nature of SFQ logic implies that the clock distribution networks (CDN) will have relatively high insertion delay which makes efficient clock gating difficult. 
To address these issues, we develop a custom circuit solution that can enable high throughput communication between clock domains. In particular, we propose a custom CDN and interface for our proposed FIFO that can support bi-communication between clock domains. 

\begin{table*}
\caption{Comparison of the JJ area and count of the proposed versus baseline FIFOs. The number of instances of each element are written in brackets after its total area.}
\centering
\tabcolsep=0.11cm
\begin{tabular}{ | c | c | c | c | c | c | c | c | c | c |c|}
\hline
\textbf{Number of} & \multicolumn{4}{c}{\textbf{JJ-area of 32-bit}} & \multicolumn{5}{|c|}{\textbf{JJ-area of 32-bit}} & \textbf{\% Increase in}  \\
\textbf{Stages} & \multicolumn{4}{c}{\textbf{previous FIFO \cite{gdatta_tas} ($\mu m^{2}$)}} & \multicolumn{5}{|c|}{\textbf{proposed FIFO ($\mu m^{2}$)}} & \textbf{JJ-area} \\
\cline{2-10}
& C-elements & DFFs & Splitters & \textbf{Total} & C-elements & DFFs & Splitters & OR gates & \textbf{Total} & \textbf{Total}\\
\hline
3 & {7(1)} & {897.68(98)} & {431.64(99)} & {1336.32} & {14(2)} & {906.84(99)} & {436(100)} & 10.56 (1) & {1367.4} & {2.32} \\
\hline
{5} & {21(3)} & {1483.92(162)} & {719.4(165)}  & {2224.32} & {28(4)} & {1493.08(163)} & {723.76(166)}  & 10.56(1) & {2255.4} & {1.39} \\
\hline
{10} & {56(8)} & {2949.52(322)} & {1438.8(330)} & {4444.32} & {63(9)} & {2958.68(323)} & {1443.16(331)} & 10.56(1)  & {4475.4} & {0.69}\\
\hline
\end{tabular}
\vspace{-0.5cm}
\label{tab:jj_area}
\end{table*}
\subsection{Design Details}

Our design employs two FIFOs, and hence has two valid and ready signals, which together with the proposed interface, enable  bi-directional data transfer.
Each interface receives data from one FIFO and writes to the other. The valid signal generated by the first clock domain controls the data transfer to the FIFO clocked by the second clock domain. Once this second FIFO successfully receives a token from the interface, its' ready port generates a pulse, which along with the valid pulse from the first clock domain controls the transfer of the next token between the interface and FIFO$_1$. Note that this valid pulse will be generated when the other interface similarly communicates with FIFO$_1$. 

Our proposal is explained below in detail and illustrated in Fig. \ref{fig:loop_circuit}, where each NDRO is denoted as ND. We also note that logic gates that are connected in a local loop in the clock domain datapath, can be grouped and clocked together using one of the locally generated clocks.
Note that the NDs and the DFFs in the control FSMs of both the datapaths are preset to logic ‘$1$' assuming each pipeline stage holds valid data. Also, note that the clock, reset, and I/O pins are at the top, bottom, and either sides of the ND respectively in Fig. \ref{fig:loop_circuit}. 

The steps involved in the bi-directional communication are described below and illustrated in 
in Fig. \ref{fig:timing_diagram}.



1) When the first pulse of $CLK_1$ arrives, every ND will produce an output pulse, then the preset DFFs will each generate an output pulse that clocks their respective pipeline datpath stages and, concurrently, reset their associated NDs. For example, the first ND will be reset by the pulse on the net labelled 'e'. Also, note that this same sequence of actions happens in parallel in domain $CLK_2$.

2) When the second $CLK_1$ pulse arrives, only the leftmost ND directly connected to FIFO$_1$ will produce an output pulse because the others have been reset in the previous cycle. The last DFF in the FSM control, denoted FSM$_1$ in Fig. \ref{fig:loop_circuit}, in addition to resetting the ND receiving the signal ‘$5$' in its' reset port, also sets the ND tied to signal ‘$6$'. The latter ND is clocked by CLK$_1$, and its' output pulse is connected to the write port of FIFO$_2$, signaling it is ready to transfer data. 

3) Note that until FIFO$_2$ signals that it is ready to accept new data by sending out a ready pulse, the data in domain $CLK_1$ will remain stalled in the interface, even if FIFO$_1$ tries to push more tokens into the pipeline. This is ensured by the reset operation of the ND by the signal ‘$5$'. 

4) Once the ready pulse is produced, 
the last ND of in the top row of FSM$_1$ will be set by signal ‘$9$', while the ND below it will be reset by ‘$8$'. In the next cycle, the output pulse of the top ND will set the ND to its' left by the signal ‘$2$' and will permit any stalled CLK$_1$ token to enter FIFO$_2$.  

5) This process continues until the ND receiving the first signal ‘c' is set, which will eventually clock the read port of the FIFO$_1$. Concurrently, after the data enters FIFO$_2$, it will be transferred to the datapath stages of domain $CLK_2$. The valid signal will drive the FIFO$_2$ read port, repeating the above process.
\subsection{Timing Constraints}

To ensure we can write data into the FIFO at least every other clock cycle, 
the clock period (both $CLK_1$ and $CLK_2$) has to be larger than
\begin{equation}\label{eq:write}T_{cr}=T_{n}+2T_{s}+T_{d}+T_{se}
\end{equation}
which is the critical path delay between the ND receiving the ready pulse and ND driving the write port (see Fig. \ref{fig:loop_circuit}), where $T_{n}$, $T_{s}$, $T_{d}$, and $T_{se}$ denote the NDRO clock-to-Q delay, splitter delay, DFF clock-to-Q delay, and DFF setup time respectively. Hence, if $F_{clk} \leq \frac{1}{T_{cr}}$, the FIFO operates correctly. The above expression evaluates to ${\sim}\textbf{33}$ GHz in SFQ5ee for 32-bit data. 
The clock frequency also needs to satisfy  
\begin{equation}\label{eq:read}
F_{clk}\leq \frac{1}{{3{T_{s}}+T_{d}+T_{se}}}
\end{equation}
to ensure the data is transferred correctly to the read side with a valid pulse. The denominator in the right hand side of the above inequality is the critical path delay between the read port and the first DFF in the control path of the read side, which is ${\sim}\textbf{40}$ GHz in SFQ5ee process. Note that both the above critical paths are independent of the datapath bit-width. 

\begin{table}
\caption{BER of the proposed two-flop (clocked logic) synchronizer at different SFQ clock frequencies}
\begin{center}
\begin{threeparttable}
\begin{tabular}{|c|c|c|}
\hline
\textbf{Clock freq.} & \textbf{Write side } & \textbf{Read side}  \\
\textbf{(GHz)} & \textbf{(BER)} & \textbf{(BER)} \\
\hline
$20$  & $>0.36*10^{-21}$\tnote{*} & $>0.36*10^{-21}$\tnote{*}  \\
\hline
$25$ & $0.22*10^{-19}$  & $0.18*10^{-12}$  \\
\hline
$30$ & $0.128*10^{-15}$ &$0.26*10^{-6}$\\
\hline
\end{tabular}
\begin{tablenotes}
\item[*] beyond the numerical precision supported by JSIM.
\end{tablenotes}
\end{threeparttable}
\label{tab:ber_values}
\end{center}
\vspace{-0.8cm}
\end{table}

\section{Simulation Setup \& Results}\label{sec:simulation_setup}



To verify our proposed circuits, we designed our logic cell library in the MIT LL SFQ5ee process with Stewart McCumber parameter \cite{likharev1991} equal to 2. The JJs used are superconductor-insulator-superconductor (SIS) Nb/AlOx-Al/Nb junctions, with critical current density of $100$ $\mu$A/$\mu$m$^{2}$ and diameter of $700$ nm. 
First, we evaluate our proposed synchronizer explained in Section \ref{sec:proposed_design} for a range of operating frequencies. We compute the BER of our synchronizer for data transfer between the two asynchronous clock domains as described in \cite{gdatta_tas}.  The results,  shown in Table \ref{tab:ber_values},
show that the proposed synchronizer leads to a BER of ${\sim}0.128{*}10^{-15}$ in the write side at $30$ GHz clock frequency, which may be sufficient for most applications. Note that the read side exhibits lower BER compared to the write side because it has a splitter between the two interfacing DFFs. Moreover, the area overhead of our proposed design compared to the baseline is shown in Table \ref{tab:jj_area}. Our proposed synchronizer incurs an average area overhead of $1.39\%$ compared to the previous design \cite{gdatta_tas}, for 5 stages.  

In order to evaluate our proposal for bi-directional communication, we designed a 8-stage 32-bit wide circular shift register. We used two FIFOs to transfer the data from each set of four DFFs in the shift register as shown in Fig. \ref{fig:loop_circuit}. 
Our design not only provides extremely low BER mentioned above and satisfies
the timing constraints described in Section \ref{sec:proposed_interface}-B.

\section{Conclusions}\label{sec:conc}

We present a FIFO synchronizer, compatible with the deep gate-level pipelining observed in SFQ, which can support bi-directional data transfer between two asynchronous clock domains. We further propose an associated robust interface which can handshake with our FIFO that supports high-frequency clocks.
Our proposal demonstrates low BER and scales to large-scale SFQ designs with increased flexibility and negligible area overhead compared to previous designs.



\bibliographystyle{IEEEtran}
\bibliography{IEEEabrv,bibliography}

\end{document}